\documentclass[american, 12pt, a4paper, titlepage]{article}
\usepackage[T1]{fontenc}
\usepackage[latin1]{inputenc}
\usepackage{lmodern}
\usepackage{graphicx}
\usepackage{booktabs}
\usepackage[chatter]{rotating}
\usepackage{fancyvrb}
\usepackage{amssymb,amsmath}
\usepackage{float}
\usepackage{tikz}
\usetikzlibrary{arrows,positioning,shadows,shapes}
\usepackage{subfigure}
\usepackage{stmaryrd}
\usepackage{cite}
\usepackage{xcolor}
\usepackage{amsfonts}
\usepackage{url}
\usepackage{pdfpages}
\usepackage{colortbl}
\usepackage{xcolor}
\usepackage{multirow}
\usepackage{textcomp}
\usepackage{hanging}
\usepackage{ulem}
\usepackage{booktabs}
\usepackage{algorithm,algorithmic}

\begin{document}

\title{{\bf NesPrInDT}: \\{\bf Nes}ted undersampling in {\bf PrInDT}}

\author{
Claus Weihs\\
TU Dortmund University\\
Faculty of Statistics\\
{\normalsize claus.weihs@tu-dortmund.de}
\and
Sarah Buschfeld\\
TU Dortmund University\\
Faculty of Cultural Studies\\
{\normalsize sarah.buschfeld@tu-dortmund.de}
}
\date{\vspace{1cm}\today}

\begin{titlepage}

\maketitle

\end{titlepage}

\noindent{\bf Abstract}\vspace{0.2cm}\\
In this paper, we extend our PrInDT method (Weihs, Buschfeld 2021) towards additional undersampling of one of the predictors. This helps us to handle multiple unbalanced data sets, i.e.\ data sets that are not only unbalanced with respect to the class variable but also in one of the predictor variables. Beyond the advantages of such an approach, our study reveals that the balanced accuracy in the full data set can be much lower than in the predictor undersamples. We discuss potential reasons for this problem and draw methodological conclusions for linguistic studies.

\section{Introduction}

In the last three decades, an interesting sociolinguistic trend has emerged in a number of former colonies of the British Empire in which English was traditionally spoken as a second language (L2). In particular in Southeast Asia and Africa, rising numbers of speakers speak English as a first language (L1) from birth, most prominently in Singapore. L1 Singapore English (SingE) has its origin in an L2 variety that emerged as the result of language contact and general mechanisms of language acquisition in adult learners during British colonization. It therefore shows a number of linguistic characteristics that are different from what we find in the typical standard L1 varieties, British and American English (Buschfeld 2020). The current study compares L1 SingE child data that was collected in 2014 to SingE adult data from the 1990s. We aim to investigate whether quantitative differences in the use of non-standard features exist between the child and adult varieties. From a linguistic perspective, this is interesting because such an investigation allows for conclusions about potential quantitative language change occurring in Singapore.

We investigate the use of zero and realized subject pronouns in two corpora of child and adult SingE by means of the PrInDT approach developed by Weihs and Buschfeld (2021). Additionally, we introduce a first important extension to the PrInDT approach, namely nested undersampling in PrInDT -- {\bf NesPrInDT}. PrInDT is a decision tree-based method in the software R (R Core Team 2019) that combines prediction and interpretation for linguistic studies (and potentially other areas of application).

Section~\ref{sec:2} presents the data used for the present study and Section~\ref{sec:3} introduces our methodological approach. In Section~\ref{sec:4}, we present the results of the study. We ultimately discuss the linguistic as well as statistical implications in Section~\ref{sec:5}. Section~\ref{sec:6} offers a brief conclusion.

\section{Data}\label{sec:2}

In the present study, we look into the realization of subject pronouns which are variably realized or not (zero) in both L2 and L1 SingE. This is illustrated in Example 1 on the basis of the child data (further examples, see below).
\begin{itemize}
\itemsep0pt
\item[1.] Researcher: And this guy is picking some flowers for his mummy and\\
Child: Now {\bf \o} [HE] is done picking the flower.
\end{itemize}
The data for this study come from the spoken part of the Singapore component of the International Corpus of English (ICE-Singapore) and parts of the ChEsS Corpus (Children's English in Singapore; Buschfeld 2020). The ChEsS data were collected by means of different elicitation methods, ranging from free interaction to linguistic experiments. The ChEsS sub-corpus for this study amounts to 36,000 words.
The ICE-data come from the 90 transcripts of approximately 2,000 words each in the spoken component > dialogues > private > face-to-face-conversations section. They comprise an overall of 202,000 words from 254 adults (ages 18 and over).

The data from both sets were manually coded for the realization of subject pronouns (\textit{I, you, he, she, it, we, you, they}). All in all, 3,225 tokens were extracted from the child corpus (2,899 realized and 326 zero) and 17,325 tokens from the adult corpus (16,543 realized and 782 zero). This does not only constitute a high imbalance in token frequencies between the small and the large classes but also between the child and adult tokens. We explicate how we meet this problem in Section~\ref{sec:3}.

The aim of our analysis is to find prediction rules for the use of subject pronouns (realized vs. zero) by means of extra- and intralinguistic variables. Table~\ref{tab:1} summarizes the variables, their levels, and the abbreviations used in the analysis. Mean length of utterance (MLU) is a measure that determines the syntactic complexity of utterances young children make and thus their grammatical development.
The differences between the pronoun types relate to their semantic reference and are exemplified in Examples 2 through 6:
\begin{itemize}
\itemsep0pt
\item[2.] referential (\textit{I, you, he, she, we, you, they}):\\
R (Researcher): What do you do with your friends?\\
Child: Sometimes {\bf \o} [WE] play some fun things.
\item[3.] referential \textit{it}:\\
R: Ah, there is the CD, right!\\
Child: {\bf \o} [IT] Was in here.
\item[4.] contextual referential \textit{it}:\\
R: So what do you want to do next? [\ldots] It's up to you.\\
Child: {\bf \o} [IT] Is up to you.
\item[5.] expletive \textit{it}:\\
Child: I think in MH370, I think they can find because {\bf \o} [IT] is easy to go there [\ldots].
\item[6.] demonstrative:\\
R: Boy boy, what is this?\\
Child: {\bf \o} [THIS] Is cheese
\end{itemize}

\noindent Note that the levels for the adult speakers (from the 1990s ICE-Sing) are one-way manifestations since the corpus lacks the kind of sociolinguistic information available for ChEsS.

\begin{table}[h]
\vspace{-0.2cm}
\centering
\caption{Overview of variables used}
\label{tab:1}
\footnotesize
\vspace{0.3cm}
    \begin{tabular}{l|l|l|l|}
    {\bf variable type} & {\bf variable} & {\bf levels} & {\bf abbreviation} \\
    \toprule
    dependent & class & realized, zero & \\
    \midrule
    independent & pronoun type & referential (refer), & PRN\_TYPE \\
    & & demonstrative (dem), & \\
    & & expletive \textit{it} (it\_ex), & \\
    & & referential \textit{it} (it\_ref), & \\
    & & contextual referential \textit{it} (it\_con)& \\
    \cmidrule{2-4}
    & mean length & 2, 3, adult & MLU  \\
    & of utterance & &  \\
    \cmidrule{2-4}
    & (ethnic) group & Chinese children (C), & ETHN\_GROUP  \\
    & & Indian children (I), &  \\
    & & adults (n\_a) & \\
    \cmidrule{2-4}
    & age & in months of the individual child;& AGE \\
    & & for all adults = 216 months; & \\
    & & numeric & \\
  \end{tabular}
\vspace{-0.3cm}
\end{table}
\normalsize

\section{Statistical Modeling}\label{sec:3}

Our methodological approach is an extension of the PrInDT approach developed by Weihs and Buschfeld (2021), which is based on decision trees (ctrees, Hothorn et al. 2006). Let us briefly recall the basic terminology and underlying ideas of this approach, in particular the notions of resampling, undersampling, prediction, and balanced accuracy. In PrInDT, predictive power is identified as the most important criterion to assess decision trees. In order to assess predictive power in data with a high imbalance between the small and the large classes, a simple way of undersampling is used as a resampling procedure. We repeatedly employ the full sample of the smaller class together with a small percentage of the larger class as the training set of the decision tree. The predictive power is assessed by means of the balanced accuracy of the two classes on the full sample. This way, accuracies of decision trees from different undersamples can be easily compared. As an additional criterion, the interpretability of the trees is taken into account, i.e.\ uninterpretable combinations of variable values are automatically excluded. In the present example, no interpretability restrictions apply. For further details on the PrInDT approach, the interested reader is referred to Weihs and Buschfeld (2021).

Since the current data set is characterized not only by a high imbalance in the class variable but also in one of the independent variables (SPEAKER), we employ a two-step procedure of nested undersampling to level the imbalance between the two levels of SPEAKER, namely child and adult. In a first step, we randomly undersample the adult data to match the size of the child corpus at ten repetitions (outer loop). In a second step, the large class of the class variable (realized) is undersampled (inner loop). For each undersample of the adult data together with the full sample of the child data, we apply the PrInDT procedure to undersample the large class of the class variable. In our example, the large class was randomly undersampled to 6\% at 999 repetitions. The algorithmic structure of our nested undersampling approach is illustrated in Algorithm~\ref{alg:1}.

\begin{algorithm}
\caption{Nested undersampling}
\label{alg:1}
\small
\begin{algorithmic}
\REQUIRE full set of child and adult data
 \STATE {\bf outer loop} (on full data):
 \FOR{repetitions in outer loop (= 10)}
   \STATE under$_{out}$ = undersample of adult data + full sample of child data;
   \STATE {\bf inner loop} (on under$_{out}$):
   \FOR{repetitions in inner loop (= 999)}
     \STATE under$_{in}$ = undersample of large class + full sample of small class;
     \STATE construct a decision tree based on under$_{in}$
   \ENDFOR
   \STATE identify best trees from inner loop corresp.\ to balanced accuracies on under$_{out}$
 \ENDFOR
 \STATE compute balanced accuracies of best trees in outer loop on full sample
\end{algorithmic}
\end{algorithm}

All results were created by means of the software R (R Core Team 2019). The decision trees were generated by the R-package `party'. We use conditional inference trees (ctrees) with the default significance level 0.05. The following results are generated by means of the R-function {\bf NesPrInDT}, that constitutes an extension of the PrInDT function.\footnote{The source code of both functions can be requested from the first author by e-mail.}

In the following, we report the best tree from nested undersampling as well as the best ensembles that consist of three trees each. Since in our example no interpretability restrictions apply, we consider the balanced accuracies of all 10 x 999 trees for identifying the best trees. The three best trees were determined for each of the ten SPEAKER undersamples. For all 30 trees, we measure the balanced accuracies on the full sample. This procedure provides the best overall tree for the full sample.

\section{Results}\label{sec:4}

\subsection{The best tree}\label{subsec:4.1}

\begin{figure}[H]
\centering
\includegraphics[angle=90,scale=0.465]{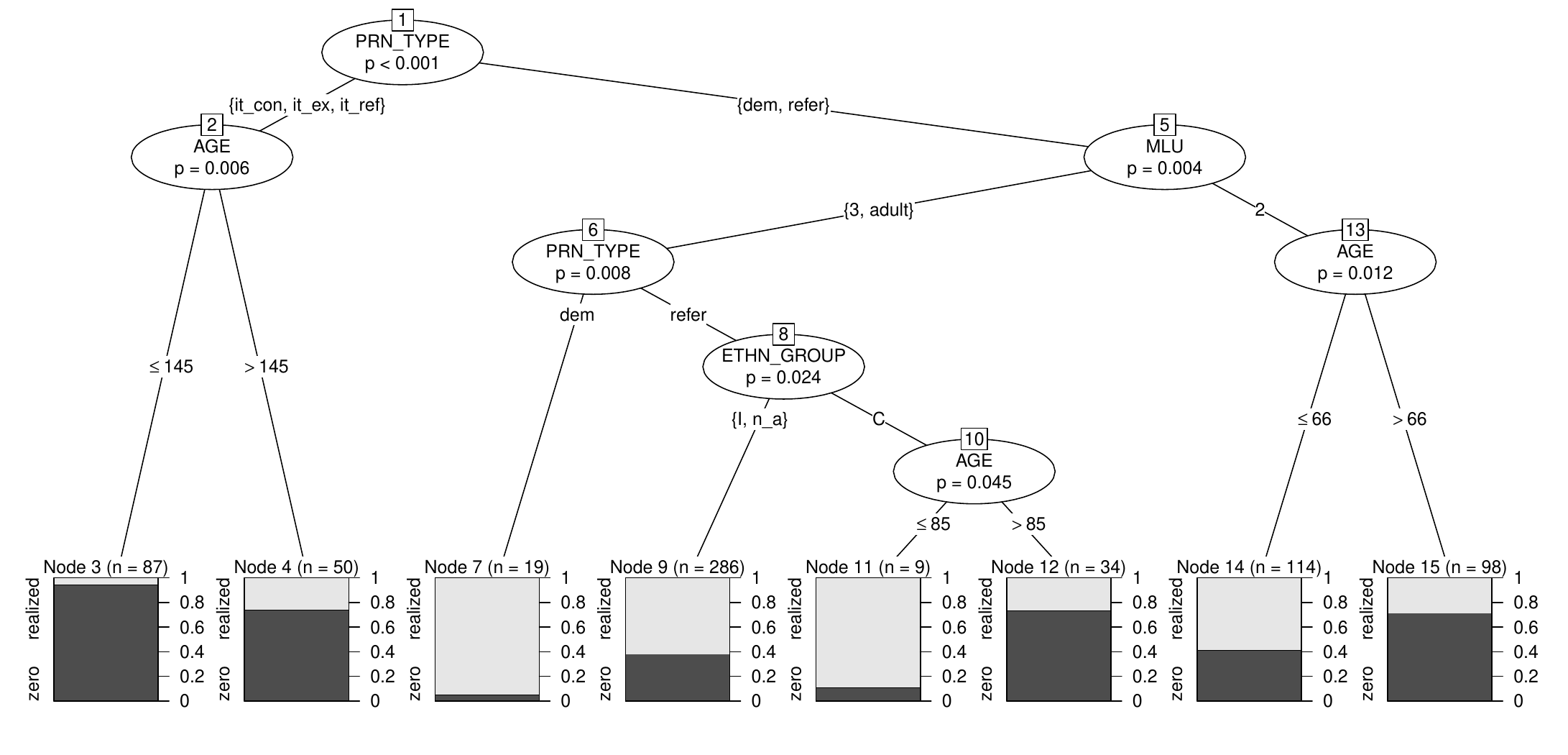}\\
\vspace{-0.2cm}
\caption{Zero vs. realized pronoun in the child and adult corpora.}
\label{fig:1}
\end{figure}

The tree with the best balanced accuracy (0.6635) based on the SPEAKER undersamples is identical to the tree with the best balanced accuracy (0.5741) based on the full sample. As Figure~\ref{fig:1} illustrates, pronoun type is the strongest predictor for subject pronoun realization. Node 1 splits the data into the three types of \textit{it} on the one hand and referential pronouns and demonstratives on the other. The concrete realization of \textit{it} further depends on the group of speakers. Node 2 splits the data into the children ($\leq$ 145 months) and the adults (> 145). For the former, the rate of zero subjects is much higher than for the adults (nodes 3 and 4). Figure~\ref{fig:1} shows that the frequency of zero subject pronouns is generally highest for \textit{it} (all three types) as used by the Singaporean children.

For referential and demonstrative pronouns, the data are further split by MLU (node 5). Surprisingly, the older children (MLU 3; children 7 years and older) are grouped with the adults (for a discussion, see Section~\ref{sec:5}). MLU group 2 is further split by AGE at 66 months (node 13). It is also surprising that the older children are the ones that realize zero pronouns significantly more frequently than the younger ones since normally the younger children go through a zero-subject phase in language acquisition. However, this finding can be explained on the basis of the data set since the MLU group 2 children older than 66 months all speak Chinese and thus a null subject language as their other L1.

The group of adults and MLU 3 children is again split by pronoun type (node 6). For demonstrative pronouns as used by this group the rate of zero pronouns is lowest in the overall tree. Group 3 children and adults hardly ever use zero pronouns of this type (node 7). Whether the group 3 children and adults make use of zero referential pronouns further depends on the speaker group again (node 8). Here, the Chinese children use significantly higher rates of zero subjects than the Indian children and the adults (with the exception of one child, see split 10). This finding does not necessarily contradict the observation that important differences in zero realization exist between the child and 1990s adult data as is discussed in Section~\ref{sec:5}.

\subsection{Balanced accuracies}\label{subsec:4.2}

The histograms of the balanced accuracies for the SPEAKER undersamples and the full sample are shown in Figures 2 and 3, respectively. Obviously, the undersample accuracies are generally higher than the full sample accuracies. One reason for this might be that the share of the smaller class is systematically higher in the SPEAKER undersamples than in the full sample, since all tokens of the children are included in the SPEAKER undersamples and, as we have seen above, zero pronouns are much more prominent for the children than for the adults. This leads to a better representation of the smaller class and therefore higher balanced accuracies in the SPEAKER undersamples than in the full sample. In fact, the best tree has an accuracy of 75\% for the larger class and 58\% for the smaller class on the SPEAKER undersample, but 84\% accuracy for the larger class and only 31\% accuracy for the smaller class on the full sample.

\begin{figure}[H]
\centering
\vspace{-0.9cm}
\includegraphics[width=0.8\textwidth]{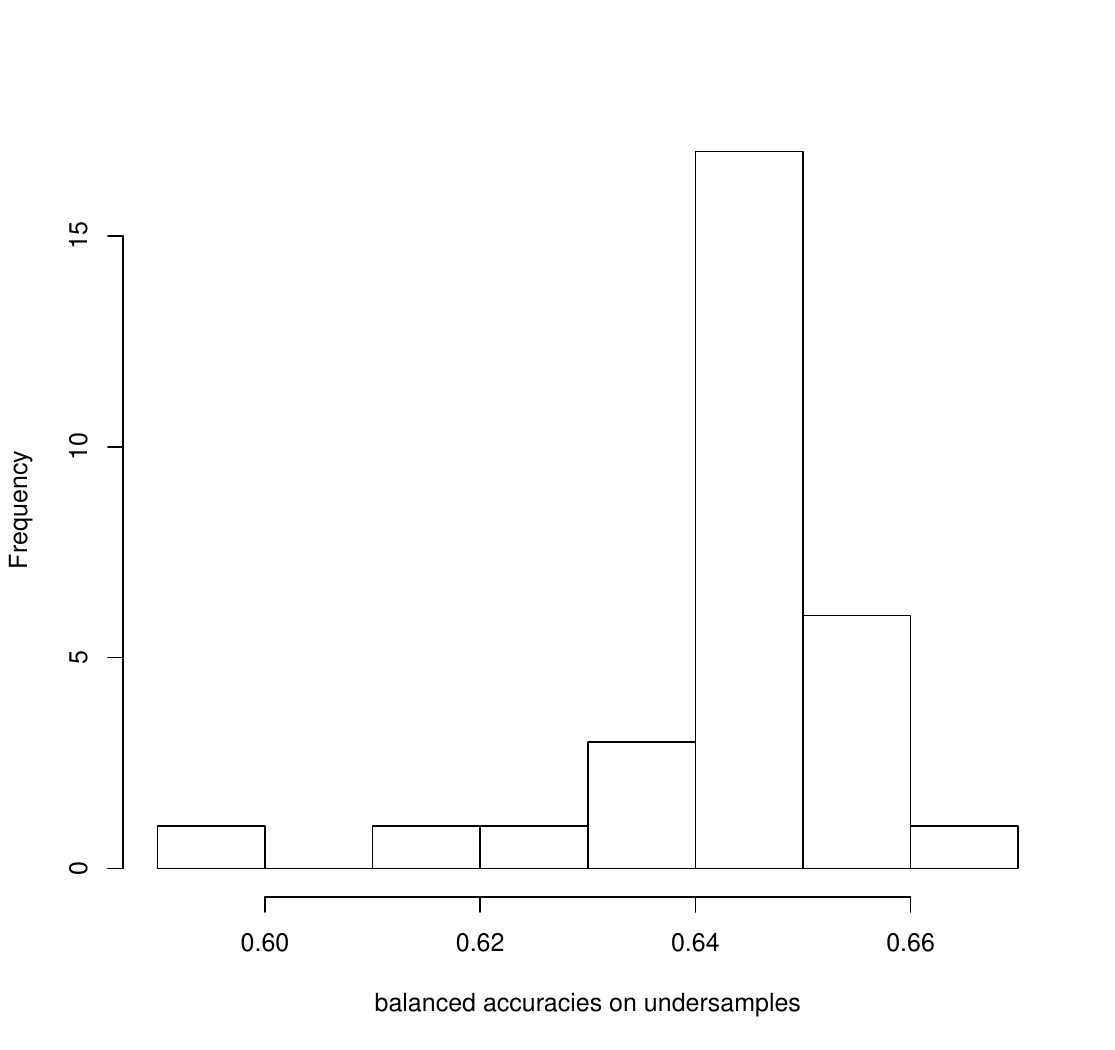}\\
\vspace{-0.3cm}
\caption{Balanced accuracies on SPEAKER subsamples.}
\label{fig:2}
\end{figure}

\begin{figure}[H]
\centering
\vspace{-0.9cm}
\includegraphics[width=0.77\textwidth]{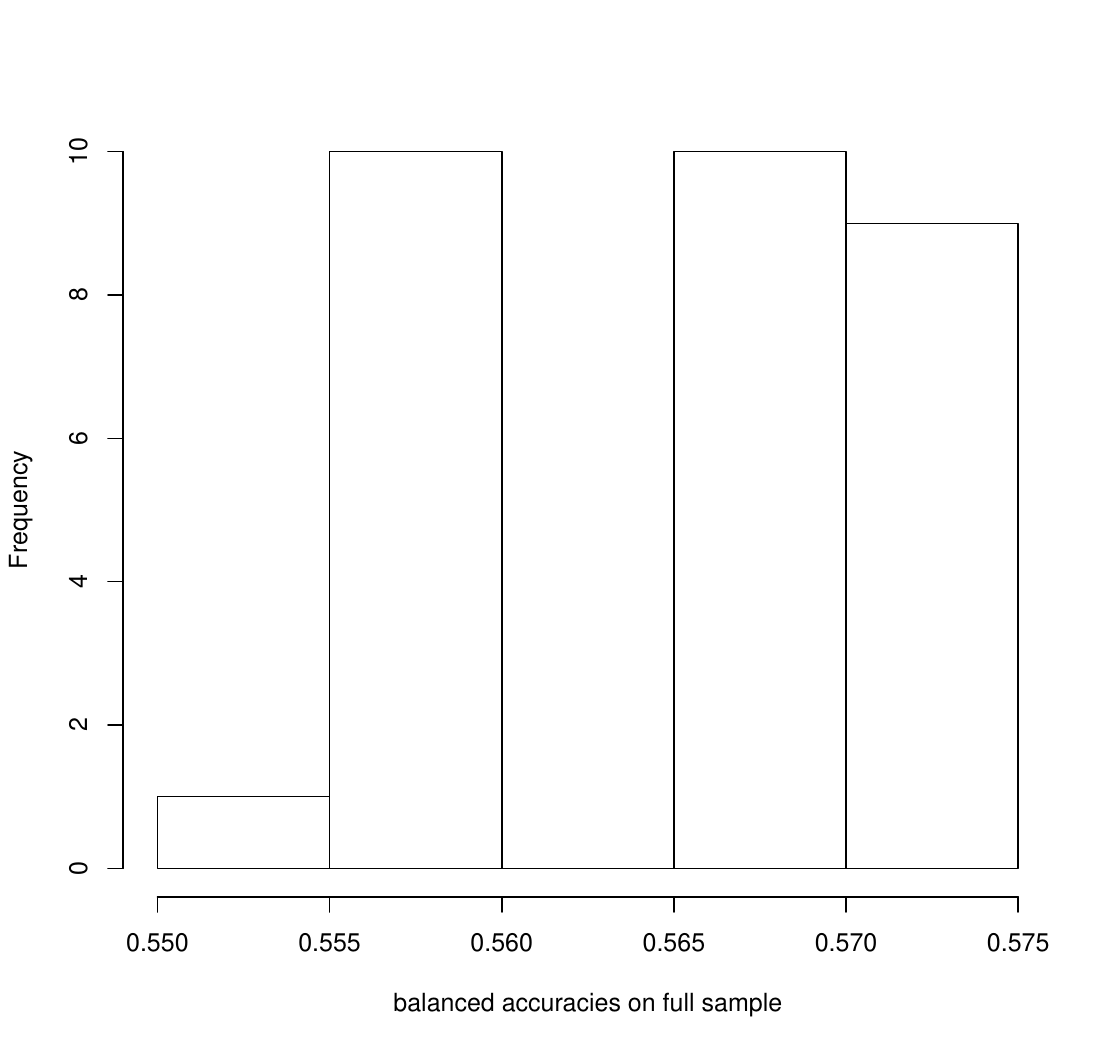}\\
\vspace{-0.3cm}
\caption{Balanced accuracies on full samples.}
\label{fig:3}
\end{figure}

Moreover, the adults appear to be very heterogeneous in their pronoun realization. To show this, we divided the adult data into eight parts according to their order of appearance in the data set. The balanced accuracies of the eight trees that comprise all tokens of the children and one-eighth of the tokens of the adults, respectively, show very different values: one tree has a balanced accuracy of 0.5, six trees of 0.566, and one tree of 0.805. In the part of the data set with the highest accuracy, the smaller class appeared much more often than in the other parts. On the one hand, this improves the accuracy. On the other hand, this illustrates the heterogeneity of the adults' sample. Unfortunately, no additional information is available which helps explain the heterogeneity in the adult sample.

\subsection{Ensembles}\label{subsec:4.2}

In a next step, we turn towards ensembles. We restrict our study to creating and evaluating ensembles from the three best trees, following two different strategies. First of all, we build ensembles of the three best trees from each of the ten SPEAKER undersamples. In the present example, the best ensemble comes with a balanced accuracy of 0.6595. Second, we compute the balanced accuracies of all 30 trees from the SPEAKER undersamples on the full sample and build an ensemble from the three best trees. This ensemble has a balanced accuracy of 0.5731. Note that both balanced accuracies are close to the best balanced accuracies of the SPEAKER undersamples and the full sample, respectively.

\section{Discussion}\label{sec:5}

In the following, we first discuss what the findings presented in Section~\ref{sec:4} reveal about our linguistic objectives: our study aims to investigate whether quantitative differences in the use of non-standard features exist between the child and the adult data sets and what the findings suggest about potential language change in SingE. Most importantly in this respect, we found statistically significant differences between the children and adults (cf. node 2, Figure~\ref{fig:1}). However, we have also seen that the split between the two groups exclusively relates to pronoun type \textit{it}, i.e.\ quantitative change can be detected for a subset of the pronouns only. Furthermore, our ctree has revealed another interesting, age-related split in node 5. For referential and demonstrative pronouns, the oldest child cohort (7$+$ years) behaves similarly to the adults and significantly different from the other two child groups. This reinforces the observation that a quantitative change occurs for the pronouns \textit{it} only but not for the rest of the pronouns. The latter finding might be interpreted as indicative of an early-stage age-grading effect. It has often been observed that children behave differently from adults but that under increased formal pressure towards more standardized language use (e.g. through the school system), they often change towards the more standardized adult speech forms later in life.

The reasons why the significant difference is restricted to pronoun type \textit{it} is most likely intralinguistic in nature. It comes as no surprise that the rate of zero subject pronoun realization is highest for the semantically empty `dummy \textit{it}' (expletive \textit{it}) since these forms have no semantic referent. What is interesting, though, is that the two other types of \textit{it}, and in particular the referential \textit{it}, behave so similarly to the expletive \textit{it}, even though they have a clear semantic referent (see Examples 3 and 4 in Section~\ref{sec:2}). The similarities in behavior must therefore be due to the phonological similarities between the three types (see also Buschfeld 2020, 182) and phonological assimilation/elision in forms such as \textit{it's}, which are frequent in spoken discourse.

From a statistical perspective, we have seen that the balanced accuracies for the best tree (and also the ensembles) vary by nearly 10\% depending on whether they are measured on the SPEAKER undersample or the full data set. We identified two potential explanations for this. First of all, the smaller class was better represented in the SPEAKER undersamples than in the full data sets. Second, we ascribed the difference in accuracy rates to a potentially strong heterogeneity in the adult data set. The data set includes adults aged 18 years and older but does not provide any information on the exact age of the participants. It thus represents an extremely broad range of ages without allowing for a stratification of the data sample accordingly; the same holds for other sociolinguistic variables such as ethnicity or gender that might help to explain variation in the data sample. To verify the heterogeneity in the adult data, we probed the data set by means of controlled resampling. To this end, we divided the adult data into eight parts according to their order of appearance in the data set and measured the balanced accuracies of all eight samples, each together with the full child data set. This procedure has indeed revealed that very high differences exist between the balanced accuracies of the eight trees; they range from 0.5 to a maximum of 0.805.

Not being able to account for this heterogeneity by means of sociolinguistic variables is, of course, problematic when it comes to discussing potential language change. Increasing numbers of L1 speakers of SingE have been reported since the 1980s (e.g.\ Kwan-Terry 1986). This suggests that change from L2 to L1 has already started in the adult population of the 1990s and we do not know whether any of the speakers in the 1990s corpus are L1 speakers of English, which is another important piece of information the ICE-Sing does not offer. In general, we cannot expect that the change we observe has taken place solely in the approximately 20 years between our two data sets.

\section{Conclusion}\label{sec:6}

In this paper, we have introduced an extended version of the PrInDT approach, {\bf NesPrInDT}, that employs nested undersampling to account for multiple unbalanced data sets.
From the above observations, we can draw some interesting linguistic-metho\-dological and statistical conclusions. First of all, we have shown that nested undersampling is a useful method to approach multiple unbalanced data sets, i.e.\ data sets that are not only unbalanced between the large and small classes but also when it comes to the distribution of tokens for one of the predictor variables (here SPEAKER). Multiple unbalanced data sets are often found in linguistic studies and certainly also in other subject areas. In principle, nested undersampling could be generalized towards even more than two variables. However, a disadvantage of nested undersampling and other forms of resampling lies in the fact that only parts of the data set are considered for modeling. We have seen that, in particular, if the larger set is very heterogeneous this may pose a problem. Nested undersampling is thus mainly suitable for homogeneous data sets. Still, we have presented a way to at least partly meet this issue. Controlled resampling of the adult data set has helped us to interpret the high difference between the balanced accuracies calculated for the SPEAKER undersamples and for the full data set.

From a linguistic perspective, the study has shed some first light into potential quantitative language change in SingE. From a methodological point of view, it clearly shows that it is very important to either personally know the data set, e.g.\ through having collected the data yourself, or to have the full set of demographic information on each speaker available (see also Kirk, Nelson 2018, 704). Many of the more recent corpora very fortunately provide such information. Still, we would like to conclude that corpus size cannot be the most important criterion for data selection (as at least implicitly inherent in the current big-data mania). Contextualization is crucial for interpretation!

\section{References}
\begin{hangparas}{.25in}{1}
Buschfeld, S. (2020), `Children's English in Singapore: Acquisition, Properties, and Use', Routledge.

Hothorn, T., Hornik, K., Zeileis, A. 2006. Unbiased recursive partitioning: a conditional inference framework. J. Comput. Graph. Stat. 15, 651--674.

International Corpus of English: \url{http://ice-corpora.net/ice/index.html}.

Kirk, J., Nelson, G. 2018. The International Corpus of English project: A progress report. World Englishes 37, 697--716.

Kwan-Terry, A. 1986. The acquisition of word order in English and Cantonese interrogative sentences: a Singapore case study. RELC Journal 17(1), 14--39.

R Core Team 2019, `R: A language and environment for statistical
  computing'. R Foundation for Statistical Computing. \url{https://www.R-project.org/}.

Weihs, C., Buschfeld, S. 2021. Combining Prediction and Interpretation in Decision Trees (PrInDT) - a Linguistic Example. arXiv: \url{http://arxiv.org/abs/2103.02336}.

\end{hangparas}

\end{document}